# Constructing feature variation coefficients to evaluate feature learning capabilities of convolutional layers in steganographic detection algorithms of spatial domain

Ru Zhang (1), Sheng Zou (1), Jianyi Liu (1), Bingjie Lin (2) and Dazhuang Liu (1)

((1)Beijing University of Posts and Telecommunications, Beijing 100876, China,(2)State Grid Information & Telecommunication Branch, Beijing 100761, China)

Correspondence should be addressed to Jianyi Liu; liujy@bupt.edu.cn

## Abstract

Traditional steganalysis methods generally include two steps: feature extraction and classification.A variety of steganalysis algorithms based on CNN (Convolutional Neural Network) have appeared in recent years. Among them, the convolutional layer of the CNN model is usually used to extract steganographic features, and the fully connected layer is used for classification. Because the effectiveness of feature extraction seriously influences the accuracy of classification, designers generally improve the accuracy of steganographic detection by improving the convolutional layer. For example, common optimizing methods in convolutional layer include the improvement of convolution kernel, activation functions, pooling functions, network structures, etc. However, due to the complexity and unexplainability of convolutional layers, it is difficult to quantitatively analyze and compare the effectiveness of feature extraction. Therefore, this paper proposes the variation coefficient to evaluate the feature learning ability of convolutional layers. We select four typical image steganalysis models based CNN in spatial domain, such as Ye-Net, Yedroudj-Net, Zhu-Net, and SR-Net as use cases, and verify the validity of the variation coefficient through experiments. Moreover, according to the variation coefficient , a features modification layer is used to optimize the features before the fully connected layer of the CNN model , and the experimental results show that the detection accuracy of the four algorithms were improved differently.

## 1. Introduction

Since the 1990s, the information hiding detection technology has evolved from the detection of specific algorithms to the blind detection of feature extraction combined with learning classification. So far, various algorithms have reached hundreds of types. Such algorithms generally include two steps: feature extraction and classification. Many algorithms use integrated classifiers (Ensemble Classifiers[1], EC) or Support Vector Machines[2] (SVM) for classification, such as image space Steganalysis algorithms SPAM[3], SRM[4], PSRM[5],TLBP[6], etc. Because the accuracy of features greatly affects the quality of steganographic detection algorithms, in the past algorithm designers spent a lot of time manually constructing efficient steganographic feature vectors. With the development of Deep Learning[7] (DL) and Graphic Processing Units[8](GPUs), some experts began to use deep learning algorithms such as convolutional neural networks to automatically extract steganographic features, which helps accurately extract features and reduce workload. Among them, Convolutional Neural Networks (CNN[9]) performs well in steganographic detection in the spatial domain, as it is able to automatically extract complex statistical dependencies from images more accurately, improving the accuracy of steganographic detection[10]. Qian-Net[11], Xu-Net[12], Ye-Net[13], etc. are good examples of such algorithms.

The detection object of the steganalysis algorithm is secret information hidden in the high-frequency data of the image, where the low-frequency data of the image may become a kind of interference. Therefore, when constructing features, generally the image is first passed through a high-pass filter to obtain multiple residual images. Statistical modeling is then performed in a single residual image. Projected histograms, co-occurrence matrices, etc. of the correlation between neighboring pixels or coefficients are used as features, and finally the features on different residual maps are combined to obtain the final feature set.

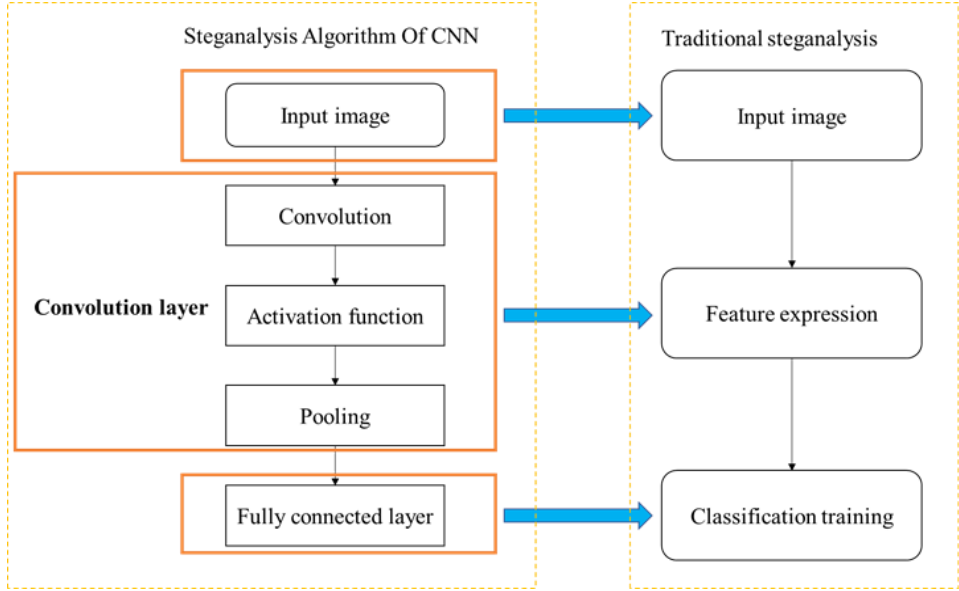

Figure 1: Functional comparison of CNN steganalysis algorithm and traditional steganalysis algorithm.

In Figure 1, the convolution kernel is functionally equivalent to the high-pass filter of the traditional method, and it can obtain the residual characteristics of the image. The activation function implements non-linear transformation and the pooling layer implements feature reduction, which is equivalent to traditional truncation and quantization. The fully connected layer is functionally equivalent to a traditional classifier.

Feature extraction greatly affects the classification effect. Bad features result in scattered similar samples, which does not help classification. Good features make similar samples relatively concentrated, making it easier to classify. Therefore, in order to improve a CNN-based steganalysis algorithm, designers often improve the convolution layer to better extract features, as shown in Tables 1 and 2. Because the process of extracting feature vectors based on CNN (convolutional neural network) steganalysis algorithms is relatively complicated, it is difficult for us to directly and quantitatively evaluate the effectiveness of the features extracted by the convolutional layers of various steganography detection algorithms. Zhong[14] et al. integrated the embedding probability estimation into the CNN to extract steganalytic features. Ren[15] et al. proposed learning selection channels for image steganalysis in spatial domain.Chen[16] et al. made the first attempt to automate the neural network architecture design for detection of global manipulations.Zeng[17] et al. proposed a novel residual CNN model with stego-signal diffusion for halftone image steganalysis.Wang[18] et al. addresses the feature representation for steganalysis of spatial steganography. Wu[19] et al. proposed a filter subset selection method to develop a well-designed pre-processing layer for CNN-based steganalysis framework. Wu[20] et al. explored an important technique in CNN models, the batch normalization (BN).Su[21] et al. proposed an ensemble classification strategy. Wang et al. proposed joint multi-domain feature learning for image steganalysis based on CNN[22].

In this paper, the validity of the features is reflected by the degree of aggregation of similar samples, and a variation coefficient index is proposed to quantitatively measure the feature learning ability of the steganalysis algorithm based on CNN. After experimental tests, the algorithm feature learning ability ranking of the coefficient of variation measurement is basically consistent with the algorithm's steganographic detection accuracy rate. Subsequently, we modify the features automatically extracted by the algorithm based on the coefficient of variation, which improved the detection accuracy of various steganographic detection algorithms to varying degrees.

The rest of this article is organized as follows. In Section 2, various CNN-based image steganalysis algorithms are reviewed. Our proposed coefficient of variation index is described in Section 3. Followed by Section 4, a feature selection method based on the coefficient of variation is proposed. The experimental results and analysis of the coefficient of variation and feature selection method are given in Section 5. Finally, concluding observations are given in section 6.

## 2. Steganography detection algorithm based on CNN

In 2014, Tan[23] et al. proposed a steganography detection scheme based on CNN. The scheme uses a four-layer neural network: three convolutional layers and a fully connected layer. It use three modes for experiments. Mode 1: Randomly initialize the first layer of convolution kernels; Mode 2: Use the KV[24] kernel to initialize the first layer of convolution kernels; Mode 3: Use the auto-encoder for pre-training based on scheme 2. The performance of the models in the three modes is compared with SPAM and SRM. As a result, it is found that the model in mode 3 outperforms SPAM but not SRM. Though not perfect, it tells us that there is great potential for steganography detection using CNN networks. Since then, scholars have paid more attention to the application of CNN in steganalysis. The focus is mainly on using CNN for feature learning, which then helps design and optimize the model structure[25].

We can refer to Table1 for the characteristics and applicability of the spatial steganalysis scheme based on deep learning, and refer to Table 2 for the characteristics and applicability of the JPEG domain steganalysis scheme based on deep learning(This article only discusses the steganalysis of the image. Audio[26] and others are not covered in this article). Comparing Table 1 and Table 2, we can see that, unlike image recognition based on deep learning, steganalysis pays more attention to image noise residuals than image content. Correspondingly, the steganographic analysis scheme based on deep learning has the following unique features:

(1) The goal of convolution layer improvement is to help calculate the image noise residual. For example, some schemes use a fixed high-pass filter placed in front of the convolution layer as a pre-processing layer, and do not update the convolution kernel; some schemes initialize the first convolution layer, and the convolution kernel can learn to update.
(2) The improvement of the pooling layer focuses on retaining more noise residual features. For example, maximum pooling will cause more residual loss, so some steganalysis schemes use average pooling. In order to prevent average pooling from averaging adjacent samples in the feature map and suppressing hidden signals, some schemes even disable the pooling layer or the pooling layer of the previous convolution layer.

Table 1: Comparison of spatial steganography detection schemes based on deep learning in recent years.

| Literature name | Algorithm characteristics |
| --- | --- |
| Stacked Convolutional Auto-Encoders for Steganalysis of Digital Images[23] | Network structure: 9-layer, three-stage CNN<br>Input: Fixed size, fixed embedding algorithm, fixed embedding rate |
| Feature Learning for Steganalysis Using Convolutional Neural Networks[27] | Network structure: 5-layer CNN network with image preprocessing<br>Input: Fixed size, fixed embedding algorithm, fixed embedding rate |
| Deep learning is a good steganalysis tool when embedding key is reused for different images, even if there is a cover source-mismatch[28] | Network structure: 2-layer CNN network with pre-processing l<br>Input: Fixed size, fixed embedding algorithm, fixed embedding rate<br>Convolution: using large-scale convolution kernels, multiple neurons |
| Ensemble of CNNs for Steganalysis: an Empirical Study[29] | Network structure: 6-layer CNN network with image preprocessing<br>Input: Fixed size, fixed embedding algorithm, fixed embedding rate |
| Learning and Transferring Representations for Image Steganalysis Using Convolutional Neural Network[30] | Network structure: 5-layer CNN network with image preprocessing and using transferring learning<br>Input: Fixed size, fixed embedding algorithm, fixed embedding rate |
| Deep Learning Hierarchical Representations for Image Steganalysis[13] | Network structure: 10-layer CNN network with image preprocessing, increasing the selection channel function according to the algorithm's probability chart<br>Input: Fixed size, fixed embedding algorithm, fixed embedding rate<br>Convolution: The first layer is initialized using the SRM core and updated<br>Activation function: Using the new hybrid activation group<br>Pooling: The first three layers are disabled for pooling |
| Yedroudj-Net: An Efficient CNN for Spatial Steganalysis[31] | Network structure: 5-layer CNN network with image preprocessing<br>Input: Fixed size, fixed embedding algorithm, fixed embedding rate<br>Activation function: The first two layers using trunc<br>Pooling: The first three layers are disabled for pooling |
| Steganalyzing Images of Arbitrary Size with CNNs[32] | Network structure: 8-layer CNN network with image preprocessing<br>Input: Fixed embedding algorithm, fixed embedding rate |

| | Convolution: Add a statistical moment extraction layer between the convolution layer and the fully connected layer |
|---|---|
| Adversarial Examples Against Deep Neural Network based Steganalysis[33] | Network structure: Multiple trained model cascades, using regression<br>Input: Fixed size, fixed embedding algorithm |
| ReST-Net: Diverse Activation Modules and Parallel Subnets-Based CNN for Spatial Image Steganalysis[34] | Network structure: Using three already trained subnets<br>Input: Fixed size, fixed embedding algorithm, fixed embedding rate<br>Convolution: Separate convolutions<br>Pooling: different activation functions |
| Deep Residual Network for Steganalysis of Digital Images[35] | Network structure: 12-layer CNN network, without image preprocessing, with residual mechanism<br>Input: Fixed size, fixed embedding algorithm, fixed embedding rate<br>Pooling: The first two layers are disabled for pooling |
| Depth-wise separable convolutions and multi-level pooling for an efficient spatial CNN-based steganalysis[36][37] | Network structure: 6-layer CNN network with image preprocessing<br>Input: Fixed embedding algorithm, fixed embedding rate<br>Convolution: Separate convolutions<br>Pooling: The first two layers and last three layer are disabled for pooling. Pyramid pooling before full connection layer. |
| A Siamese CNN for Image Steganalysis[38] | Input: Fixed embedding algorithm, fixed embedding rate<br>Pooling: Disabled for pooling. Using Global pooling after feature extraction. |

Table 2: Comparison of JPEG steganography detection schemes based on deep learning in recent years.

| Literature name | Algorithm characteristics |
|---|---|
| Deep Convolutional Neural Network to Detect J-UNIWARD[39],[40] | Network structure: Based on Siamese architecture.<br>Input: Fixed size, fixed embedding algorithm, fixed embedding rate<br>Convolution: With residual mechanism<br>Pooling: Using global pooling |
| JPEG-Phase-Aware Convolutional Neural Network for Steganalysis of JPEG Images[41] | Network structure: Utilize existing networks, Using DCT cores and catalytic cores<br>Input: Fixed size, fixed embedding algorithm, fixed embedding rate |
| Large-Scale JPEG Image Steganalysis Using Hybrid Deep-Learning Framework[42] | Network structure: 3-layer CNN network with image preprocessing(Initializing with manual)<br>Input: Fixed size, fixed embedding algorithm, fixed embedding rate<br>Convolution: With BN[43] |

| | Activation function: With ABS in the first convolution layer<br>Pooling: The first layers is disabled for pooling |
|---|---|
| Automatic steganographic distortion learning using a generative adversarial network[43] | Network structure: Using Generative Adversarial Network (GAN)<br>Input:Fixed size, fixed embedding algorithm, fixed embedding rate |
| Deep convolutional neural network-based feature extraction for steganalysis of content-adaptive JPEG steganography[44] | Network structure: Using 4 subnet with 4 kind of filters, pooling is disabled for the first two layers in subnet<br>Input:Fixed size, fixed embedding algorithm, fixed embedding rate |

Now we select four CNN-based steganographic detection algorithms with good detection rates from Tables 1 and 2, and further analyze and compare their network structure characteristics.

*2.1 a series of high pass filter initialization algorithms.* In 2017, Ye[13] et al. proposed a Ye-Net network with a good steg detection rate, as shown in Figure 2. Ye-Net has a 10-layer convolutional neural network that has three characteristics. First, it introduces prior knowledge in the field of steganalysis into the model structure, improving the accuracy of detecting adaptive steganography algorithms. Second, the first layer of the network is initialized with 30 SRM convolution kernels, and the convolution kernels can be updated iteratively during training. Last but not least, Ye-Net uses an activation function TLU[45] that is more adaptive to the steganographic noise distribution and has faster convergence speed. Experiments show that Ye-Net's detection capabilities for WOW[46], S-UNIWARD[47] and HILL[48] algorithms are significantly better than traditional detection models.

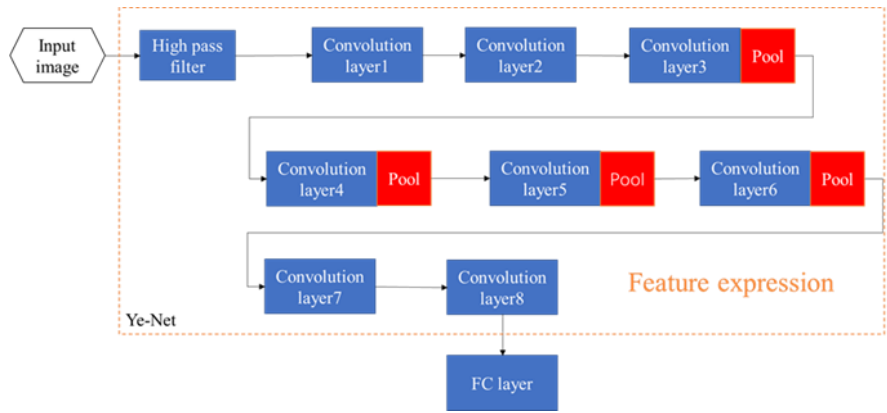

Figure 2: Ye-Net network structure.

In 2018, Yedroud[31] et al. used a fixed filter to filter the image in the preprocessing layer, and pooling was disabled in the first convolutional layer in order to prevent the hidden signals from being suppressed by averaging neighboring samples in the feature map during the average merge. A truncation activation function is added to the first two convolution layers, as shown in Figure 3. Experiment results show that Yedroud-Net performs slightly better than Ye-Net.

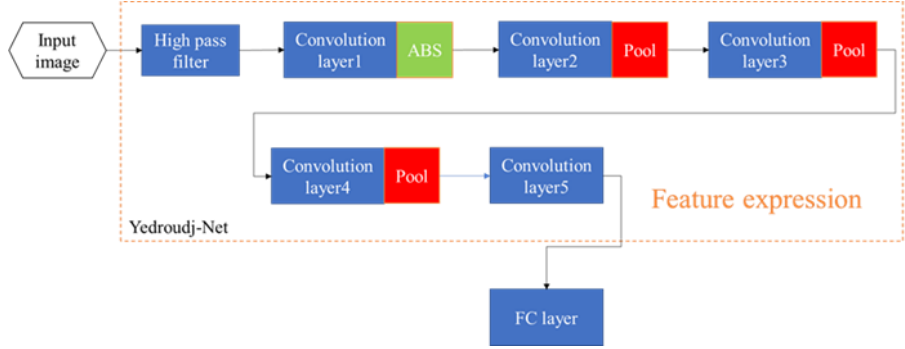

Figure 3: Yedroudj-Net network structure.

In 2020, Zhang[36] et al. proposed the ZhuNet network as shown in Figure 4. They modify the size of the convolution kernel at the preprocessing layer, and use 30 basic filters of SRM[4] to initialize the convolution kernel at the preprocessing layer to reduce the number of parameters and optimize local features, and then optimize the preprocessing layer by training convolution kernels to achieve better accuracy and accelerate network convergence. In the next layer two separable convolutional blocks[49][50] are used instead of the traditional convolutional layer. The merit of separated convolution blocks is that they can be used to extract the spatial correlation and channel correlation of residuals to increase the signal-to-noise ratio, which significantly improves algorithm accuracy. Before the fully connected layer, the spatial pyramid pool[51] is used to process the feature vectors obtained by the network and map the feature map to a fixed length; extraction was performed through a multi-stage pool to achieve processing of images of different sizes.

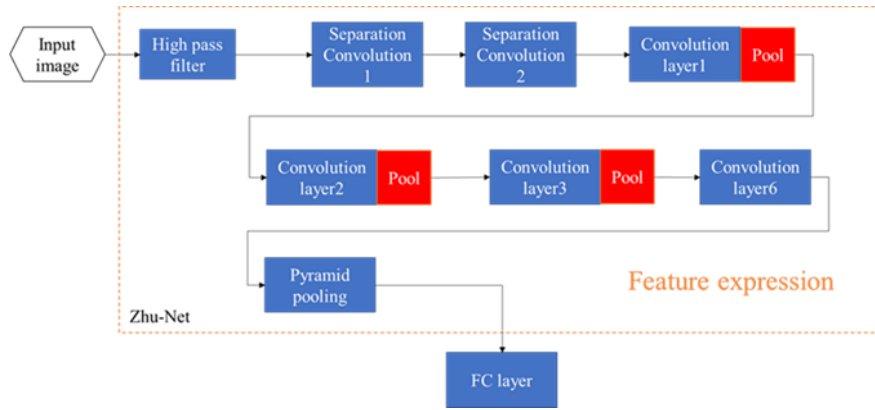

Figure 4: Zhu-Net network structure.

*2.2 SR-Net network with self-learning high-pass filter.* In 2018, Boroumand[35] et al. proposed SR-Net for detecting airspace and Jpeg domain, as shown in Figure 5. Boroumand replaces the traditional high-pass wave filter with 7 convolution layers in series connection, which is used to obtain the noise residual of the image and disable the pooling operation to prevent the hidden signal loss, so that the model can be used in the air domain and the Jpeg domain. After that, the four convolutional layers with residual mechanism are connected to extract the features. Finally, a convolutional layer is connected for dimensionality reduction and compression.

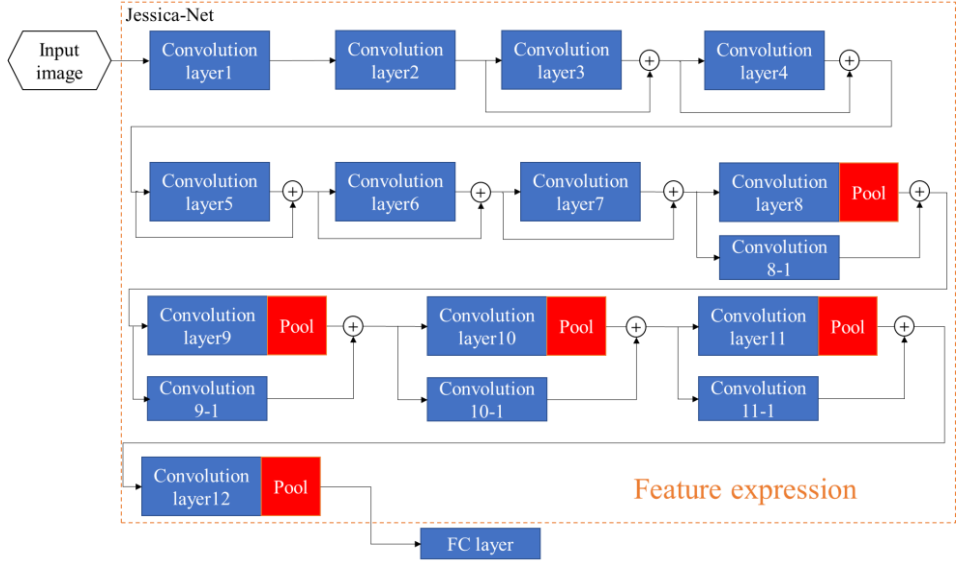

Figure 5: SR-Net network structure.

## 3. Common performance indicators

In the field of spatial domain steganalysis technology research, the traditional indicators to measure the performance of certain steganalysis algorithms are usually as follows:

*3.1 Accuracy.* The common indicator of the steganalysis algorithm. $Y$ is total number of tested samples. $X$ is total number of correctly tested samples. The detection accuracy $P$ is as follows.

$$P = \frac{X}{Y} \times 100\% \tag{1}$$

*3.2 Confidence Interval.* Assuming that there are n samples to be tested, which are randomly divided into m groups, and $Y(y_1, y_2, \ldots, y_m)$ is the number of correctly classified samples in each group, and the accuracy of each group is:

$$\delta_i = \frac{y_i}{\frac{n}{m}} \times 100\% \tag{2}$$

Therefore, the average accuracy rate is:

$$\bar{\delta} = \frac{\sum_{i=1}^{m} \delta_i}{m} \tag{3}$$

Let $z$ be the critical value constant, $\sigma$ be the average accuracy rate, and $n$ be the number of samples.

When there are enough samples to satisfy the normal distribution, the confidence interval of the model is:

$$c = 10 \times z \times \sqrt{\bar{\delta} \times (1 - \bar{\delta}) \div m} \tag{4}$$

Among them, $z$ is a critical value constant, and the constant values of 98%, 95%, and 90% confidence are shown in Table 3:

Table 3: Critical value constant z value table.

|  | Threshold constant | | |
| --- | --- | --- | --- |
|  | 98% | 95% | 90% |
| Confidence Values |  |  |  |
| Critical Value | 2.33 | 1.96 | 1.64 |

# 4. Variation coefficient used for modifying steganographic features

In this section, we classify all samples into cover class and stego class.

*4.1 Feature variation coefficients of stego class.* Because the feature maps of different algorithms have different aggregation degrees by observing, we construct the feature variation coefficient to test the aggregation degrees.

Taking the feature vector of the cover and stego samples obtained from the convolution layer as input, the feature variation coefficients of the two types of samples are constructed according to the following steps:

**S1:** Let $X_{nm}$ be the set of feature vectors of the same sample, that is,

$$\begin{pmatrix} x_{11} & \cdots & x_{1m} \\ \vdots & \ddots & \vdots \\ x_{n1} & \cdots & x_{nm} \end{pmatrix} \tag{5}$$

Among them, $n$ is the number of similar samples, $m$ is the number of dimensions of the feature vector, and is the j-th dimensional feature value of the i-th sample.

**S2:** Calculate the j-th dimensional feature mean of the feature vector $X_{nm}$ of the same sample by column:

$$\mu_j = \bar{X}_j = \frac{\sum_{i=1}^{n} x_{ij}}{n} \tag{6}$$

Calculate the j-th dimensional feature standard deviation of the feature vector of the same sample:

$$\sigma_j = \sqrt{\frac{\sum_{i=1}^{n}(x_{ij} - \bar{X}_j)}{n}} \tag{7}$$

**S3:** According to the formula of the coefficient of variation, we can obtain the j-th dimensional feature variation coefficient of the feature vector $X_{nm}$ of the same sample:

$$CV_j = \frac{\sigma_i}{\mu_i} \tag{8}$$

**S4:** Since the dimensionality of the feature vector of each CNN model is different, in order to facilitate comparison, we calculate the mean of the coefficient of variation $CV_j$ of each dimension of the same sample to obtain the coefficient of variation of the characteristics of the same sample:

$$\overline{CV} = \frac{\sum_{j=1}^{m} CV_j}{m} \tag{9}$$

So far, we have constructed the average coefficient of variation $\overline{CV}$ of the steganography detection algorithm to compare the feature learning capabilities of various steganography detection algorithms.

*4.2 Improving CNN model with feature set modification.* In this section, we consider filtering the "bad" features in the feature vector $X_{nm}$ according to the coefficient of variation $CV_j$ to further improve the classification accuracy. The new steganalysis procedure is shown in Figure 6. Training image sample set was used to pre-train a CNN model, to gain the feature sets. The variation coefficients of feature sets are calculated. We rebuild the CNN model by modifying some features with highest variation coefficients, and trained the new model. Testing sample set was classified with the rebuild model.

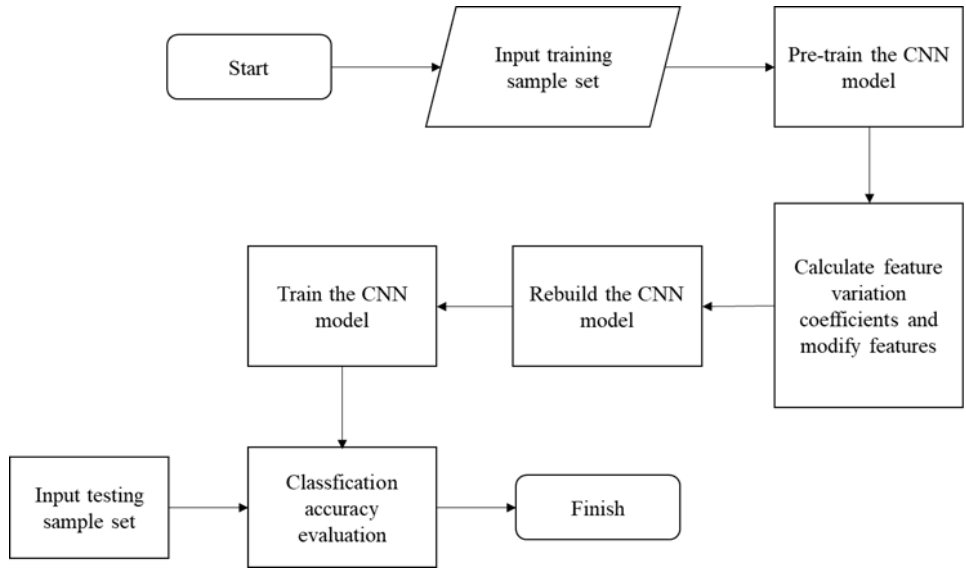

Figure 6: The new steganalysis procedure.

Rebuilding CNN model is shown as Figure 7. Features modification with variation coefficients is operated after CNN layer. Fully connection layer use optimal features for more accurate classification.

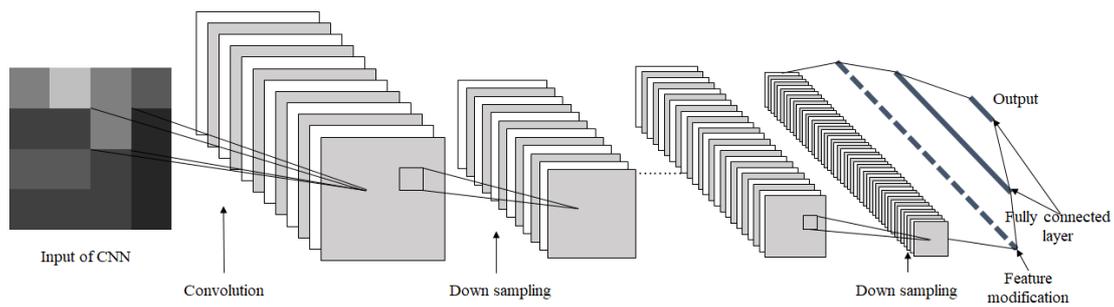

Figure 7: Rebuild CNN model.

Calculating feature variation coefficients and modifying features is shown as Figure 8, and Algorithm 1 shows the details of features modification method.

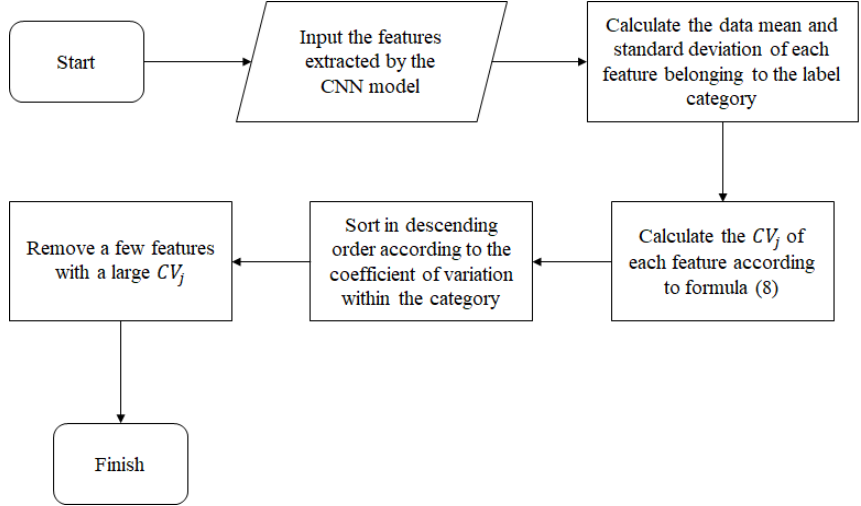

Figure 8: Rebuild CNN model.

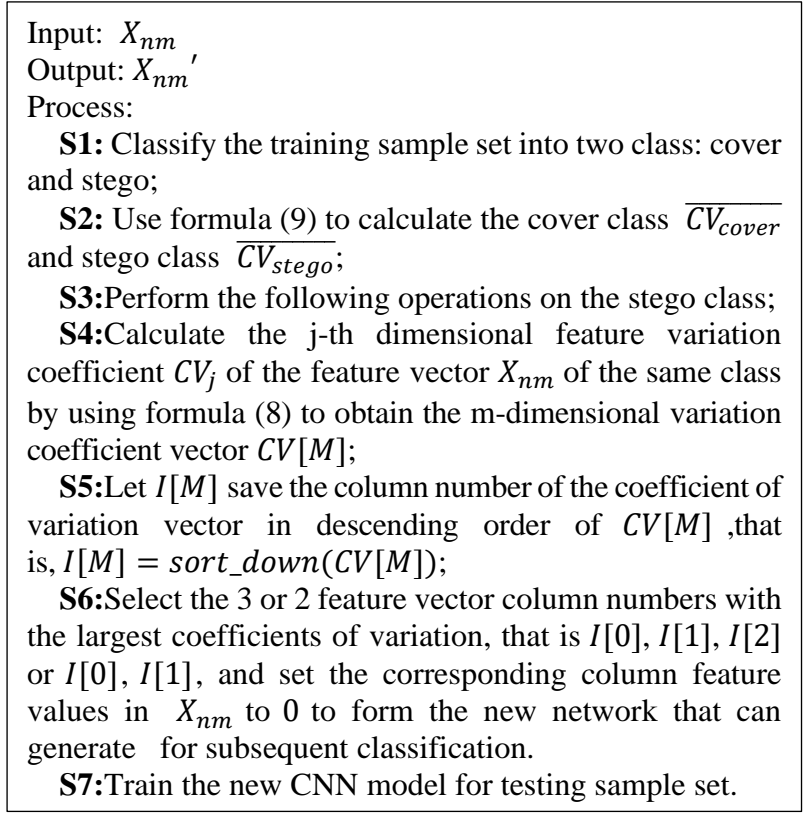

Algorithm 1: Features modification.

The main function of Method1 is to find the 3 or 2 columns of feature vectors (where 3 and 2 are empirical values) with the largest variation coefficient of the stego class, use them as anomalous feature vectors, and make them not participate in sample classification by zeroing.

It should be noted that if the number of zeroed columns is too large, the accuracy of the classification result will be reduced because the classifier feature input is too small. For the same reason, we only use the Method1 to operate on the stego class once.

## 5. Experiments and results analysis

We selected four popular steganalysis CNN models with high detection accuracy for comparison: Ye-Net, Yedroudj-Net, Zhu-Net, and SR-Net. All experiments were run on an Nvidia RTX 2080Ti GPU card.

*5.1 Data Sets.* This article uses a standard data set for comparing experiments. BOSSBase v1.01 contains 10,000 grayscale images from 7 different cameras with a size of 512 x 512 and is never compressed. Due to our GPU computing power and time constraints, we use matlab to resample the original image into a 256 x 256 pixel image for subsequent experiments.

Embedding algorithms use adaptive steganography methods in spatial domain: WOW[46]. Embeding rates are: 0.4bpp, 0.2bpp and 0.1bpp.

We randomly divide BOSSBase v1.01 images into a training set with 4,000 cover and stego image pairs, a validation set with 1,000 image pairs, and a test set containing 5,000 image pairs. No data virtual expansion was performed on the above data sets. The training, validation and test sets do not overlap.

*5.2 Hyper-parameters.* In experiments, we use Pytorch to implement the four steganalysis CNN models, and use SGD (Stochastic gradient descent algorithm) to train them. In models, momentum and weight_decay parameters are assigned 0.9 and 0.0005. Considering experimental equipment, we assign 8 to batch (eight cover/stego pairs). except for SR-Net randomly initialized (according to the original paper), models are initialized with Xavier[52] method. With these parameters, cross-entry is chosen as the loss function with 0.005 initial learning rate.

*5.3 Visualization of aggregation degree for samples class.* In general, CNN-based steganalysis convolution layers transform image samples into feature vector samples for fully connected layer classification, that is, the pictures become points in the multi-dimensional feature space, and then the points are classified and usually divided into two categories : cover class and stego class. The aggregation state of the same kind of points output by different algorithm convolution layers is different. Good features can better group similar points together. Therefore, the aggregation degree of similar points can reflect the quality of the feature. We choose the coefficient of variation as a statistic to measure the degree of discreteness of the sample in order to quantitatively measure the ability to express the steganographic detection features of the CNN network. (The coefficient of variation is a statistic used to measure the aggregation degree of each observation. It has advantages of being dimensionless, easier to calculate, and easy to compare.)

The CNN steganalysis algorithm constructs high dimensional feature sets. We use the nonlinear descending dimension algorithm, t-sne, to make any sample feature set $X\{x_1, x_2, ..., x_n\}$ descended to a visual low dimensional feature set $Y\{y_1, y_2, ..., y_n\}$。

T Distribution is used to initialize the low dimensional feature set. Let learning rate $\eta$. terative num $t$, difficulty coefficien $p$ (typically 0~50), and initial momentum $\alpha(t)$ (to avoid getting the local optical solution).

Calculate similarity of high dimensional feature sets in same type of samples (a sample class) as $p_{i|j}$ (Among them, $\sigma$ is variance).

$$p_{i|j} = \frac{\exp(-\|x_i - x_j\|^2 / 2\sigma^2)}{\sum_{k \neq l} \exp(-\|x_k - x_l\|^2 / 2\sigma^2)} \tag{10}$$

Calculate similarity of low dimensional feature sets in same type of samples (a sample class), as $q_{i|j}$.

$$q_{i|j} = \frac{(1 + \|y_i - y_j\|^2)^{-1}}{\sum_{k \neq l}(1 + \|y_k - y_l\|^2)^{-1}} \tag{11}$$

Calculate error between (before) pre-descending and (after) post-descending, and use the error for decreasing gradient optimization. Among them, both $P_i$ and $Q_j$ are distribution (distributes) of high dimensional and low dimensional feature sets.

$$C = \sum_i KL(P_i \| Q_j) = \sum_i \sum_j p_{j|i} \log \frac{j|i}{j|j} \quad (12)$$

So the gradient is,

$$\frac{\delta_C}{\delta_{y_i}} = 4 \sum_j (p_{i|j} - q_{i|j})(y_i - y_j)\left(1 + \|y_i - y_j\|^2\right)^{-1} \quad (13)$$

Renew the low dimensional feature set as follows.

$$Y^{(t)} = Y^{(t-1)} + \eta \frac{\delta_C}{\delta_Y} + \alpha(t)\left(Y^{(t-1)} - Y^{(t-2)}\right) \quad (14)$$

Figure 9 is feature maps of the convolutional layer output of the four algorithms, which has been reduced to two dimensions by t-sne for distribution visualization. It can be seen in the figure that at different embedding rates (0.4bpp, 0.2bpp, 0.1bpp), the feature points aggregation degrees in the same class extracted by different algorithms are different, and Zhu-Net's points in cover class look more clustered. The embedded algorithm is WOW.

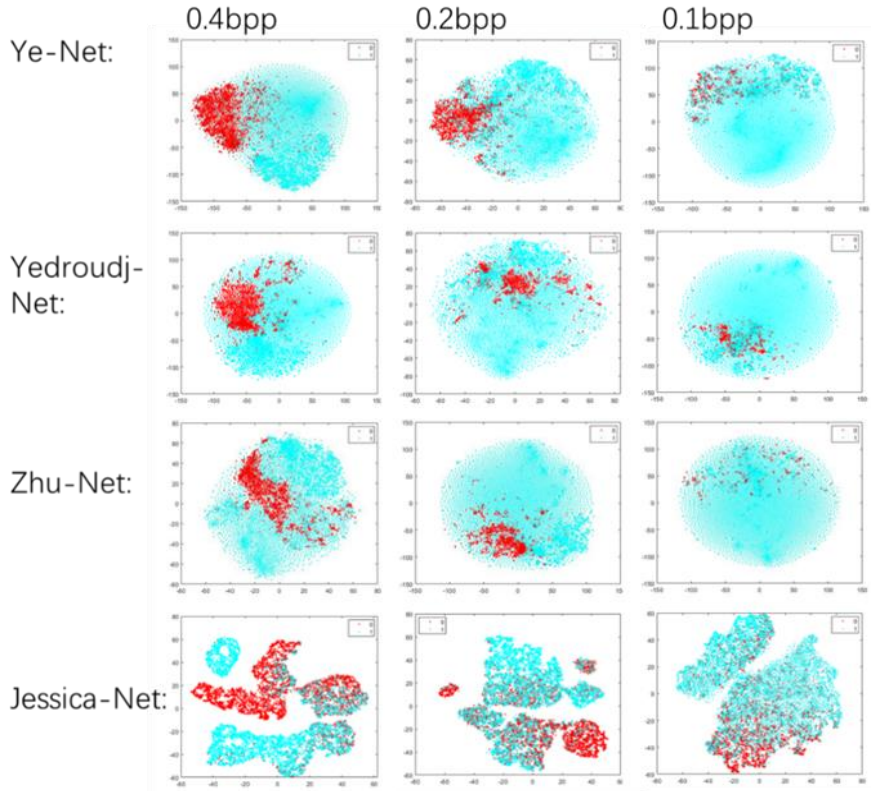

Figure 9: Visualization of the two-dimensional feature maps of the four algorithms in cover class and stego class. The red dots are the feature map of cover class, and the blue dots are the the feature map of stego class.

*5.4 Experiments.*

*5.4.1 Average variation coefficient.* First, we send the training set to each network, output a feature map in front of the fully connected layer, and store it by label, i.e., cover or stego. We then calculate the average coefficient of variation to compare the feature learning capabilities of the four algorithms. The comparison of their reciprocal and accuracy is shown in Figure 10.

Table 4: Calculate and compare the average coefficient of variation of the stego set of the four algorithms.

| Average variation coefficient | 0.4bpp | | 0.2bpp | | 0.1bpp | |
|---|---|---|---|---|---|---|
| | $\overline{CV}_{cover}$ | $\overline{CV}_{stego}$ | $\overline{CV}_{cover}$ | $\overline{CV}_{stego}$ | $\overline{CV}_{cover}$ | $\overline{CV}_{stego}$ |
| Ye-Net | 2.59 | 2.59 | 3.65 | 9.65 | 29.01 | 94.14 |
| Yedroudj-Net | 2.51 | 2.58 | 2.25 | 2.29 | 2.32 | 2.32 |
| Zhu-Net | 1.58 | 1.75 | 1.68 | 1.81 | 1.43 | 1.42 |
| SR-Net | 7.86 | 3.82 | 18.10 | 3.00 | 35.89 | 3.93 |

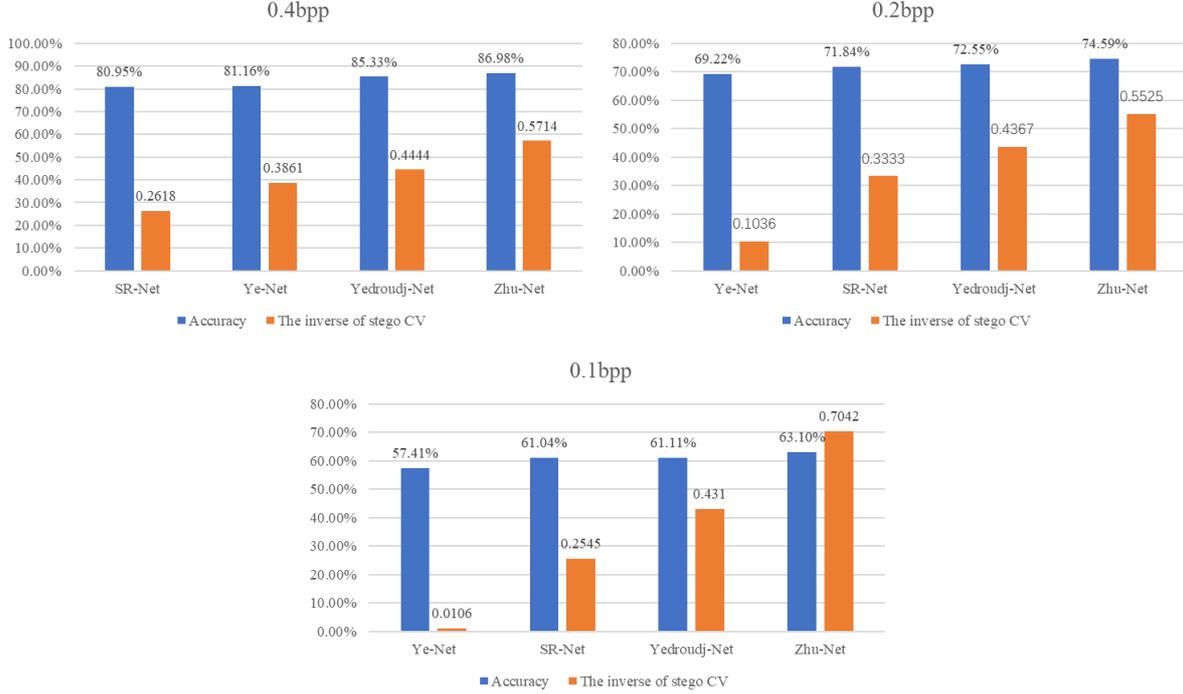

Figure 10: Accuracy and the inverse of stego CV.

As shown in Table 4, at 0.4bpp, 0.2bpp and 0.1bpp, the six terms of Zhu-Net are significantly better than the other three algorithms, followed by Yedroudj-Net, and lastly Ye-Net, which is consistent with the performance comparison of detection accuracy (see Table 6). In Zhu-Net's convolution layer, the separate convolution operation, the use of small convolution kernel operation, and the pyramid pooling operation perform better feature learning on the cover and stego classes respectively and capture the unique characteristics of various samples. Different types of samples are clustered more closely and have better feature expression capabilities. The self-learning high-pass filter algorithm of SR-Net has basically the same average coefficient of variation ranking and accuracy ranking.

As the embedding rate decreases, it may become harder for the algorithm to learn features. Both types of coefficients of variation of Ye-Net increase as the embedding rate decreases, and the coefficients of variation of cover class of SR-Net increase as the embedding rate decreases. The indexes of other algorithms are not similar. It may be that the algorithm captures different features for samples with different embedding rates. And Ye-Net and SR-Net may still learn the same features at different embedding rates.

At the same time, we also notice that the variation coefficients of the cover classes of Ye-Net, Yedroudj-Net and Zhu-Net are all lower than the variation coefficients of the stego class and the opposite of SR-Net. This phenomenon is also consistent with the false positive rate of SR-Net being higher than the false negative rate. It can be seen that SR-Net's ability to learn the features of the stego class is stronger than the cover class.

*5.4.2 Confidence interval.* In order to prove that the algorithm with a small confidence interval radius has a higher accuracy rate, we perform another group experiment.

We randomly divide the test set into 100 groups with 50 pairs of images in each group. Since the number of data sets exceeds 30, the sample can be considered to be in a normal distribution. So we use the confidence interval radius to measure the performance of the model. We test each group of images, calculate the average accuracy, and calculate the radius of the normal sub-arrangement confidence interval according to formula (4). Table 5 shows that at the embedding rate of 0.4bpp, the 98%, 95% and 90% confidence interval radii of Zhu-Net are 0.078, 0.066 and 0.055, which are 0.04, 0.03 and 0.03 lower than Yeedroudj-Net, and 0.013, 0.011, 0.009 lower than Ye-Net. Algorithms with a small confidence interval radius have better generalization and more general features. Therefore, Zhu-Net is better than Yedroudj-Net and Ye-Net. In the other two embedding rates, the measurement results of the three models are sorted the same. The specific data can be found in the table below.

Table 5: Comparison of confidence intervals for the four algorithms under three embedding rates.

| Confidence interval | 0.4bpp | | | 0.2bpp | | | 0.1bpp | | |
|---|---|---|---|---|---|---|---|---|---|
| | 98% | 95% | 90% | 98% | 95% | 90% | 98% | 95% | 90% |
| Ye-Net | 0.091 | 0.077 | 0.064 | 0.108 | 0.090 | 0.076 | 0.115 | 0.097 | 0.081 |
| Yedroudj-Net | 0.082 | 0.069 | 0.058 | 0.104 | 0.087 | 0.073 | 0.114 | 0.096 | 0.080 |
| SR-Net | 0.091 | 0.077 | 0.064 | 0.105 | 0.088 | 0.074 | 0.113 | 0.096 | 0.080 |
| Zhu-Net | 0.078 | 0.066 | 0.055 | 0.101 | 0.085 | 0.071 | 0.112 | 0.095 | 0.079 |

*5.4.3 Accuracy of rebuilt CNN model with features modification.* We filter the feature maps output by the Ye-Net, Yedroudj-Net, Zhu-Net and SR-Net algorithms with the variation coefficient. Experiments are performed at four embedding rates, and the accuracy of each algorithm is increased. Taking into account the problem of feature loss, we use method1 in section 4.2 to modify the features of the stego class. The results are shown in Table 6 to Table 7.

Table 6: Comparison of the accuracy rates before and after features modification.

| Accuracy | 0.4bpp | | 0.2bpp | | 0.1bpp | |
|---|---|---|---|---|---|---|
| | $\overline{CV_{cover}}$ | $\overline{CV_{stego}}$ | $\overline{CV_{cover}}$ | $\overline{CV_{stego}}$ | $\overline{CV_{cover}}$ | $\overline{CV_{stego}}$ |
| Ye-Net | 2.59 | 2.59 | 3.65 | 9.65 | 29.01 | 94.14 |
| Yedroudj-Net | 2.51 | 2.58 | 2.25 | 2.29 | 2.32 | 2.32 |
| Zhu-Net | 1.58 | 1.75 | 1.68 | 1.81 | 1.43 | 1.42 |
| SR-Net | 7.86 | 3.82 | 18.10 | 3.00 | 35.89 | 3.93 |

Table 7: Comparison of confidence intervals before and after features modifications.

| Confidence interval Optimized | 0.4bpp | | | 0.2bpp | | | 0.1bpp | | |
|---|---|---|---|---|---|---|---|---|---|
| | 98% | 95% | 90% | 98% | 95% | 90% | 98% | 95% | 90% |
| Ye-Net | 0.086 | 0.073 | 0.061 | 0.105 | 0.088 | 0.074 | 0.112 | 0.094 | 0.079 |
| Yedroudj-Net | 0.081 | 0.068 | 0.057 | 0.104 | 0.087 | 0.073 | 0.113 | 0.095 | 0.080 |
| SR-Net | 0.076 | 0.064 | 0.053 | 0.099 | 0.083 | 0.069 | 0.111 | 0.093 | 0.078 |
| Zhu-Net | 0.088 | 0.074 | 0.062 | 0.102 | 0.086 | 0.072 | 0.111 | 0.094 | 0.078 |

With rebuilt CNN model, the accuracy of steganalysis is increased, and the confidence interval size of each algorithm is reduced. Among the four algorithms, the performance of Ye-Net has been improved the most. The modified Ye-Net has narrowed the gap with Yedroudj-Net performance, and even surpassed Yedroudj-Net with 0.1bpp embedding rate. Under 0.1bpp, the accuracy of the original Ye-Net is 57.41%, and the adjusted rate is 63.59%, surpassing 61.60% of Yedroudj-Net, and the confidence interval is also better than that of Yedroudj-Net.

After the modification of Zhu-Net, the accuracy rate has also been improved to a certain extent, which has widened the performance gap with Yedroudj-Net.

With rebuilt model, the accuracy rate of SR-Net has been greatly improved under the two embedding rates of 0.2bpp and 0.1bpp. It can be seen that SR-Net has a unique advantage for low embedding rate steganalysis tasks.

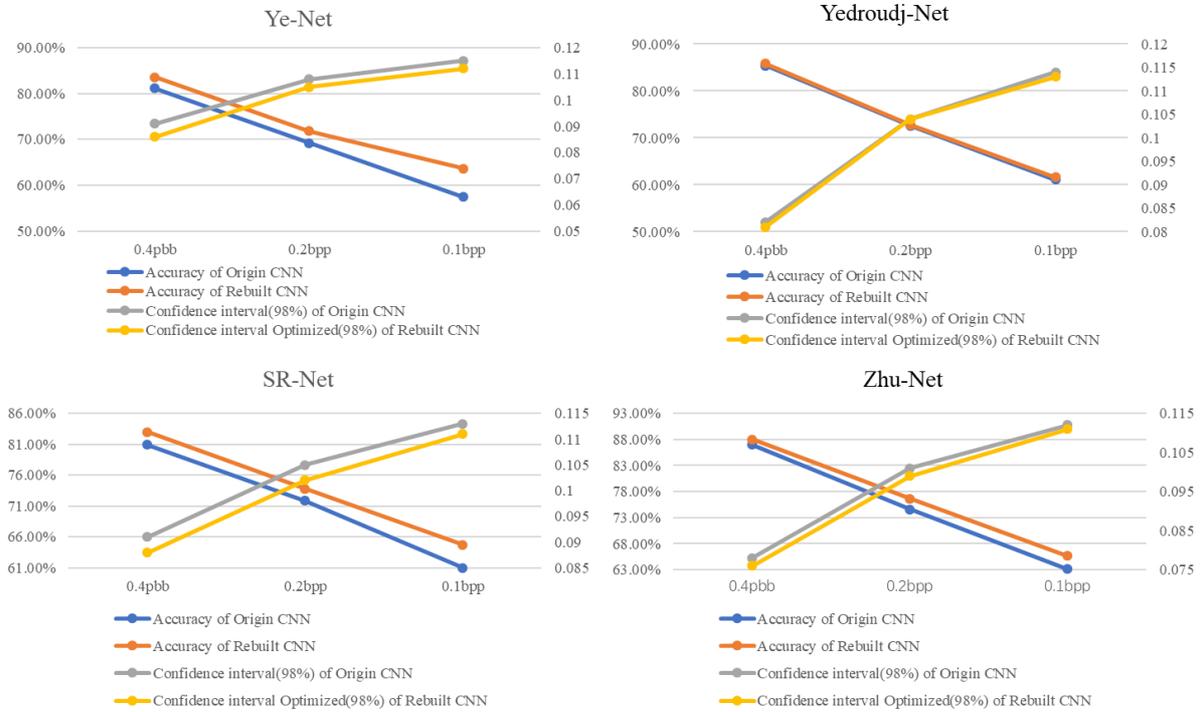

Figure 11: Accuracy and Confidence interval of four algorithms.

*5.5 Results analysis.* According to Figure 11, the accuracy of the four algorithms has been improved to various degrees. For models with poor coefficients of variation, after adjusting and optimizing the features based on the coefficients of variation, the accuracy improves more significantly. The experimental results show that the CNN steganalysis model can measure the feature expression ability of the steganalysis model from a new perspective by measuring the "feature learning ability" index, that is, the variation coefficient of the feature map and the radius of the confidence interval of the algorithm.

Among different steganalysis algorithms with initialized high-pass filters, the smaller the variation coefficient of similar samples, the more aggregated the same samples are, which is more conducive to classification, better performance, and the same trend as the accuracy rate. In contrast to the steganalysis model, the variation coefficient of the sample can still reflect the feature expression ability of the algorithm.

The confidence interval is used to measure the algorithm performance and feature expression ability, and the result is the same as the accuracy ranking.

On a standard single data set with different embedding rates, the test results of these two indicators are basically consistent with the performance ranking under the accuracy rate indicator.

In addition, the characteristics of the algorithm are optimized based on the coefficient of variation, and the accuracy of the algorithm is improved, which further proves that the coefficient of variation measures the effectiveness of the algorithm's feature expression ability.

## 6. Conclusion

The steganalysis algorithm based on CNN is a typical classification problem. Due to the complexity of the CNN structure, the main performance evaluation indicators of such algorithms are usually only detection accuracy and other auxiliary indicators. There is no quantitative evaluation index for the structure of the convolution layer. In this article, we focus on finding an index that can quantitatively measure the feature learning ability of CNN steganalysis models. After experimental tests, the variation coefficient index we constructed can be used to quantitatively evaluate the feature learning capabilities of various fully trained CNN steganalysis models. The basic principle is: use the variation coefficient to measure the aggregation degree of the feature vector output by the convolution layer. Subsequently, we can use the coefficient of variation to modify the features to aggregate the same samples more densely, which increases the accuracy of the algorithm, and further proves the effectiveness of the coefficient of variation to measure the feature learning ability of the CNN steganalysis algorithm.

In the experiment, we found that the accuracy of different algorithms has been improved to various degrees after feature modification. Next, the focus of our work will be to explore the method for optimizing convolution layer structure based on coefficient of variation, to further optimize the steganographic features of the algorithm to improve its accuracy.

## Data Availability

The data used to support the findings of this study are available from the corresponding author upon request.

## Conflicts of Interest

The authors declare that there are no conflicts of interest regarding the publication of this paper.

## Acknowledgments

This work was supported by the National Key Research and Development Program of China under Grant (2019YFB140650)，National Natural Science Foundation of China (U1936216, U1836108).

## References

[1] J. Kodovsky, J. Fridrich, and V.Holub, "Ensemble classifiers for steganalysis of digital media," IEEE Trans. Inf. Forensic Security. vol. 7, no. 2, pp. 432-444, Nov. 2011.
[2] C.-C. Chang, and C.-J. Lin, "LIBSVM: A library for support vector machines," ACM Trans. Intell. Syst. Technol. vol. 2, no. 3, pp. 1-27, May 2011. [Online]. Available: http://doi.acm.org/10.1145/1961189.1961199
[3] T. Pevny, P. Bas, and J. Fridrich, "Steganalysis by subtractive pixel adjacency matrix," IEEE Trans. Inf. Forensic Secur. vol. 5, no. 2, pp. 215-224, Mar. 2010
[4] J. Fridrich, and J. Kodovsky, "Rich models for steganalysis of digital images," IEEE Trans. Inf. Forensic Secur. vol. 7, no. 3, pp. 868-882, May 2012.
[5] V. Holub, and J. Fridrich, "Random projections of residuals for digital image steganalysis," IEEE Trans. Inf. Forensic Secur. vol. 8, no. 12, pp. 1996-2006, Oct. 2013.
[6] B. Li, Z. Li, S. Zhou, and S. Tan, X. Zhang, "New steganalytic features for spatial image steganography based on derivative filters and threshold LBP operator," IEEE Trans. Inf. Forensic Secur. vol. 13, no. 5, pp. 1242-1257, Dec. 2017.

[7] M.A. Nielsen, "Neural networks and deep learning," Determination press San Francisco, CA, 2015. pp. 875–936. [Online]. Available: https://www.sciencedirect.com/science/article/pii/B9780128015223000185

[8] R.T Soto, "Programación paralela sobre arquitecturas heterogéneas." 2016. [Online]. Available: http://www.bdigital.unal.edu.co/54267/

[9] B. Sahiner, H.-P. Chan, N. Petrick, D. Wei, M.A. Helvie, D.D. Adler, and M.M. Goodsitt "Classification of mass and normal breast tissue: a convolution neural network classifier with spatial domain and texture images," IEEE Trans. Med. Imaging vol. 15, no. 5, pp. 598-610, Oct 1996.

[10] I. Hussain, J. Zeng, and S. Tan, "A Survey on Deep Convolutional Neural Networks for Image Steganography and Steganalysis," KSII Trans. Internet Inf. Syst. vol. 14, no. 3, pp. 1228-1248, Mar. 2020.

[11] Y. Qian, J. Dong, W. Wang, and T. Tan, "Deep learning for steganalysis via convolutional neural networks," in Proc. Media Watermarking, Security, and Forensics 2015, vol. 9409, p. 94090J: International Society for Optics and Photonics. Mar. 2015.

[12] G. Xu, H.-Z. Wu, and Y.-Q. Shi, "Structural design of convolutional neural networks for steganalysis," IEEE Signal Process. Lett. vol. 23, no. 5, pp. 708-712, May 2016.

[13] J. Ye, J. Ni, and Y. Yi, "Deep learning hierarchical representations for image steganalysis," IEEE Trans. Inf. Forensic Secur. vol. 12, no. 11, pp. 2545-2557, Nov. 2017.

[14] S.-H. Zhong, Y. Wang, T. Ren, M. Zheng, Y. Liu, and G. Wu, "Steganographer Detection via Multi-Scale Embedding Probability Estimation," ACM Trans. Multimed. Comput. Commun. Appl. vol. 15, no. 4, pp. 1-23, Dec. 2019

[15] Ren, Weixiang; Zhai, Liming; Jia, Ju. Learning selection channels for image steganalysis in spatial domain. NEUROCOMPUTING. Vol.401, pp78-90. 2020.

[16] Y. Chen, Z. Wang, Z. J. Wang, and X. Kang, "Automated Design of Neural Network Architectures with Reinforcement Learning for Detection of Global Manipulations," IEEE J. Sel. Top. Signal Process.Vol.14, no. 5, pp. 997-1011, Aug. 2020.

[17] L. Zeng, W. Lu, W. Liu, and J. Chen, "Deep residual network for halftone image steganalysis with stego-signal diffusion," Signal Process. Vol. 172, p. 107576, July 2020.

[18] P. Wang, F. Liu, and C. J. S. P. Yang, "Towards feature representation for steganalysis of spatial steganography," Signal Process, vol. 169, p. 107422,Apr. 2020.

[19] L. Wu, X. Han, C. Wen, and B. Li"A Steganalysis framework based on CNN using the filter subset selection method," Multimed. Tools Appl. Vol.79, no.27-28, pp:19875-19892,2020

[20] S. Wu, S.-h. Zhong, and Y. J. I. T. o. M. Liu, "A novel convolutional neural network for image steganalysis with shared normalization," vol. 22, no. 1, pp. 256-270, Jan. 2020.

[21] A. Su and X. Zhao, "Boosting Image Steganalysis Under Universal Deep Learning Architecture Incorporating Ensemble Classification Strategy," IEEE Signal Process. Lett. vol. 26, no. 12, pp. 1852-1856, Dec. 2019.

[22] Z. Wang, M. Chen, Y. Yang, M. Lei,and Z. Dong , "Joint multi-domain feature learning for image steganalysis based on CNN," EURASIP J. Image Video Process. vol. 2020, no. 1, pp. 1-12, Jul. 2020.

[23] S. Tan and B. Li, "Stacked convolutional auto-encoders for steganalysis of digital images," in Proc. Signal and Information Processing Association Annual Summit and Conference (APSIPA), 2014 Asia-Pacific, pp. 1-4: IEEE. Dec. 2014.

[24] J. Mielikainen, "LSB matching revisited," IEEE Signal Process. Lett. vol. 13, no. 5, pp. 285-287, May 2006.

[25] T.-S. Reinel, R.-P. Raul, and I. Gustavo, "Deep learning applied to steganalysis of digital images: a systematic review," IEEE Access, vol. 7, pp. 68970-68990, May 2019.

[26] H. Tian, J. Sun, Y. Huang, T. Wang, Y. Chen, and Y. Cai, "Detecting steganography of adaptive multirate speech with unknown embedding rate," Mobile Inf. Syst., vol. 2017, no. 5418978, May 2017.

[27] Y. Qian, J. Dong, W. Wang, and T. Tan, "Feature learning for steganalysis using convolutional neural networks," Multimedia Tools Appl. vol. 77, no. 15, pp. 19633-19657, Aug. 2018.

[28] L. Pibre, J. Pasquet, D. Ienco, and M. J. E. I. Chaumont, "Deep learning is a good steganalysis tool when e mbedding key is reused for different images, even if there is a cover sourcemismatch," in Proc. Media Wat ermarking ,Secur. Forensics, San Francisco, CA, USA, Feb. 2016, pp. 14–18. [Online]. Available: http://ar xiv.org/abs/1511.04855

[29] G. Xu, H.-Z. Wu, and Y. Q. Shi, "Ensemble of CNNs for steganalysis: An empirical study," in Proc. Proceedings of the 4th ACM Workshop on Information Hiding and Multimedia Security, 2016, pp. 103-107.

[30] Y. Qian, J. Dong, W. Wang, and T. Tan, "Learning and transferring representations for image steganalysis using convolutional neural network," in Proc. 2016 IEEE international conference on image processing (ICIP), Sep. 2016, pp. 2752-2756.

[31] M. Yedroudj, F. Comby, and M. Chaumont, "Yedroudj-net: An efficient CNN for spatial steganalysis," in Proc. 2018 IEEE International Conference on Acoustics, Speech and Signal Processing (ICASSP), 2018, pp. 2092-2096: IEEE.

[32] C. F. Tsang and J. J. E. I. Fridrich, "Steganalyzing images of arbitrary size with CNNs," Electronic Imaging,

vol. 2018, no. 7, pp. 121-1-121-8, Jul. 2018.
[33] Y. Zhang, W. Zhang, K. Chen, J. Liu, Y. Liu, and N. Yu, "Adversarial examples against deep neural network based steganalysis," in Proc. Proceedings of the 6th ACM Workshop on Information Hiding and Multimedia Security, 2018, pp. 67-72.
[34] B. Li, W. Wei, A. Ferreira, and S. Tan, "ReST-Net: Diverse activation modules and parallel subnets-based CNN for spatial image steganalysis," IEEE Signal Process. Lett. vol. 25, no. 5, pp. 650-654, May 2018.
[35] M. Boroumand, M. Chen, and J. Fridrich, "Deep residual network for steganalysis of digital images," IEEE Trans. Inf. Forensic Secur. vol. 14, no. 5, pp. 1181-1193, May 2018.
[36] R. Zhang, F. Zhu, J. Liu, and G. Liu, "Depth-wise separable convolutions and multi-level pooling for an efficient spatial CNN-based steganalysis," IEEE Trans. Inf. Forensic Secur. vol. 15, pp. 1138-1150, Aug. 2019.
[37] W. Ahn, H. Jang, S.-H. Nam, I.-J. Yu, and H.-K. Lee, "Local-Source Enhanced Residual Network for Steganalysis of Digital Images," IEEE Access, vol. 8, pp. 137789-137798, Jul. 2020.
[38] W. You, H. Zhang, and X. J. Zhao,"A Siamese CNN for Image Steganalysis," IEEE Trans. Inf. Forensic Secur. vol. 16, pp. 291-306, Jul. 2020.
[39] G. Xu, "Deep convolutional neural network to detect J-UNIWARD," in Proc. Proceedings of the 5th ACM Workshop on Information Hiding and Multimedia Security, 2017, pp. 67-73.
[40] D. Hu, L. Wang, W. Jiang, S. Zheng, and B. Li, ''A novel image steganogra-phy method via deep convolutional generative adversarial networks,''IEEE Access, vol. 6, pp. 38303–38314, Jun. 2018.
[41] M. Chen, V. Sedighi, M. Boroumand, and J. Fridrich, "JPEG-phase-aware convolutional neural network for steganalysis of JPEG images," in Proc. Proceedings of the 5th ACM Workshop on Information Hiding and Multimedia Security, 2017, pp. 75-84.
[42] J. Zeng, S. Tan, B. Li, and J. Huang, ''Large-scale JPEG image steganalysis using hybrid deep-learning framework,'' IEEE Trans. Inf. Forensics Security, vol. 13, no. 5, pp. 1200–1214, May 2018.
[43] W.Tang, S.Tan, B.Li, and J.Huang,''Automatic steganographic distortion learning using a generative adversarial network,'' IEEE Signal Process. Lett., vol. 24, no. 10, pp. 1547–1551, Oct. 2017.
[44] X. Song, X. Xu, Z. Wang, Z. Zhang, and Y. Zhang, "Deep convolutional neural network-based feature extraction for steganalysis of content-adaptive JPEG steganography," J. Electron. Imaging, vol. 28, no. 5, p. 053029, Oct. 2019.
[45] I. Morishita, "Analysis of an Adaptive Threshold Logic Unit, " IEEE Trans. Comput., vol. C-19, no. 12, pp. 1181–1192, Dec. 1970.
[46] V. Holub and J. Fridrich, "Designing steganographic distortion using directional filters," in Proc. 2012 IEEE International workshop on information forensics and security (WIFS), 2012, pp. 234-239: IEEE.
[47] V. Holub, J. Fridrich, and T. Denemark, "Universal distortion function for steganography in an arbitrary domain," EURASIP Journal on Information Security, vol. 2014, no. 1, pp. 113, Jan. 2014.
[48] B. Li, M. Wang, J. Huang, and X. Li, "A new cost function for spatial image steganography," in Proc. 2014 IEEE International Conference on Image Processing (ICIP), 2014, pp. 4206-4210: IEEE.
[49] C. Szegedy, W. Liu, Y. Jia, P. Sermanet, S. Reed, D. Anguelov, D. Erhan, V. Vanhoucke, and Rabinovich, A., "Going deeper with convolutions," in Proc. Proceedings of the IEEE conference on computer vision and pattern recognition, 2015, pp. 1-9.
[50] F. Chollet, "Xception: Deep learning with depthwise separable convolutions," in Proc. Proceedings of the IEEE conference on computer vision and pattern recognition, 2017, pp. 1251-1258.
[51] K. He, X. Zhang, S. Ren, and J. Sun, "Spatial pyramid pooling in deep convolutional networks for visual recognition," IEEE Trans. Pattern Anal. Mach. Intell. vol. 37, no. 9, pp. 1904-1916, Sep. 2015.
[52] Z. Jin, Y. Yang, Y. Chen, and Y. Chen, "IAS-CNN: Image adaptive steganalysis via convolutional neural network combined with selection channel," Int. J. Distrib. Sens. Netw. vol. 16, no. 3, p. 1550147720911002, Mar. 2020.